\documentclass[a4paper,nodate]{sab} 
\usepackage{graphicx}
\usepackage{xcolor}
\usepackage{txfonts,natbib} 
\bibpunct{(}{)}{;}{a}{}{,} 
\usepackage{hyperref}
\usepackage{url}
\usepackage[T1]{fontenc}
\usepackage[utf8]{inputenc}
\usepackage{t1enc} 
\renewcommand\ackname{Acknowledgements.} 
\renewcommand\andname{\&}
\renewcommand\keywordname{Keywords.} 
\renewcommand\figureptname{Figure}
\renewcommand\tableptname{Table}

\renewcommand\noteaddname{Note added to proofs.}
\renewcommand\andname{\&}

\idline{Boletim da Sociedade Astron\^omica Brasileira, {\bf 31}, no. 1, XX-XX}{3}

\def\feii{{Fe {\sc ii}}}
\def\caii{{Ca {\sc ii}}}
\def\kms{km~s$^{-1}$}

\def\msun{M$_\odot$\/}
\def\oiii{{\sc [Oiii]}$\lambda$5007}
\def\rfe{R$_\mathrm{FeII}$}
\def\hb{H$\beta$}
\def\mbh{$M_{\rm BH}$\/}
\def\cc{cm$^{-3}$}
\def\civ{{C {\sc iv}}}
\def\nv{{N {\sc v}}}
\def\hei{{He {\sc i}}}
\def\heii{{He {\sc ii}}}
\def\ciii{{C {\sc iii}]}}
\def\lya{Ly$\alpha$}
\def\mgii{Mg {\sc ii}}
\def\oi{{O {\sc i}}}
\def\rblr{R$_{\rm BLR}$}
\def\lopt{L$_{\rm 5100}$}
\def\lbol{L$_{\rm bol}$}
\def\ledd{L$_{\rm Edd}$}
\def\mdot{$\dot{\mathcal{M}}$}

\begin{document} 
\title{Modeling the quasar spectra for super-Eddington sources} 
\subtitle{The What, the Why and the How} 

\author{S. Panda \inst{1} \and P. Marziani \inst{2}} 
\institute{Laborat\'orio Nacional de Astrof\'isica - Rua dos Estados Unidos 154, Bairro das Na\c c\~oes. CEP 37504-364, Itajub\'a, MG, Brazil. \email{spanda@lna.br}  \and INAF-Astronomical Observatory of Padova, Vicolo dell'Osservatorio, 5, 35122 Padova PD, Italy.\\ \email{paola.marziani@inaf.it}}
\date{Received } 

\Abstract {Broad lines in active galaxies are primarily emitted by photoionization processes that are driven by the incident continuum arising from a complex geometrical structure circumscribing the supermassive black hole. A model of the broad-band spectral energy distribution (SED) effective in ionizing the gas-rich broad line emitting region (BLR) is needed to understand the various radiative processes that eventually lead to the emission of emission lines from diverse physical conditions. Photoionization codes are a useful tool to investigate two aspects - the importance of the shape of the SED, and the physical conditions in the BLR. In this work, we focus on the anisotropy of continuum radiation from the very centre a direct consequence of the development of a funnel-like structure at regions very close to the black hole. Accounting for the diversity of  Type-1 active galactic nuclei (AGNs) in the context of the main sequence of quasars, permits us to locate the super Eddington sources along the sequence and constrain the physical conditions of their line-emitting BLR.}{Linhas amplas em galáxias ativas são emitidas principalmente por processos de fotoionização que são conduzidos pelo contínuo incidente decorrente da estrutura geométrica complexa que circunscreve o buraco negro supermassivo. Um modelo da distribuição de energia espectral de banda larga (SED) eficaz na ionização da região emissora de linha larga (BLR) rica em gás é necessário para entender os vários processos radiativos que eventualmente levam à emissão de linhas de emissão de diversas condições físicas. Os códigos de fotoionização são uma ferramenta útil para investigar dois aspectos - a importância da forma do SED e as condições físicas no BLR. Neste trabalho, focamos na anisotropia da radiação contínua do próprio centro, uma consequência direta do desenvolvimento de uma estrutura em forma de funil em regiões muito próximas ao buraco negro. Levar em conta a diversidade de núcleos galácticos ativos do tipo 1 (AGNs) no contexto da sequência principal de quasares, nos permite localizar as fontes super Eddington ao longo da sequência e restringir as condições físicas de seus BLR emissores de linha.}


\maketitle

\section{Introduction}

\begin{figure*}
    \centering
    \includegraphics[width=1\textwidth]{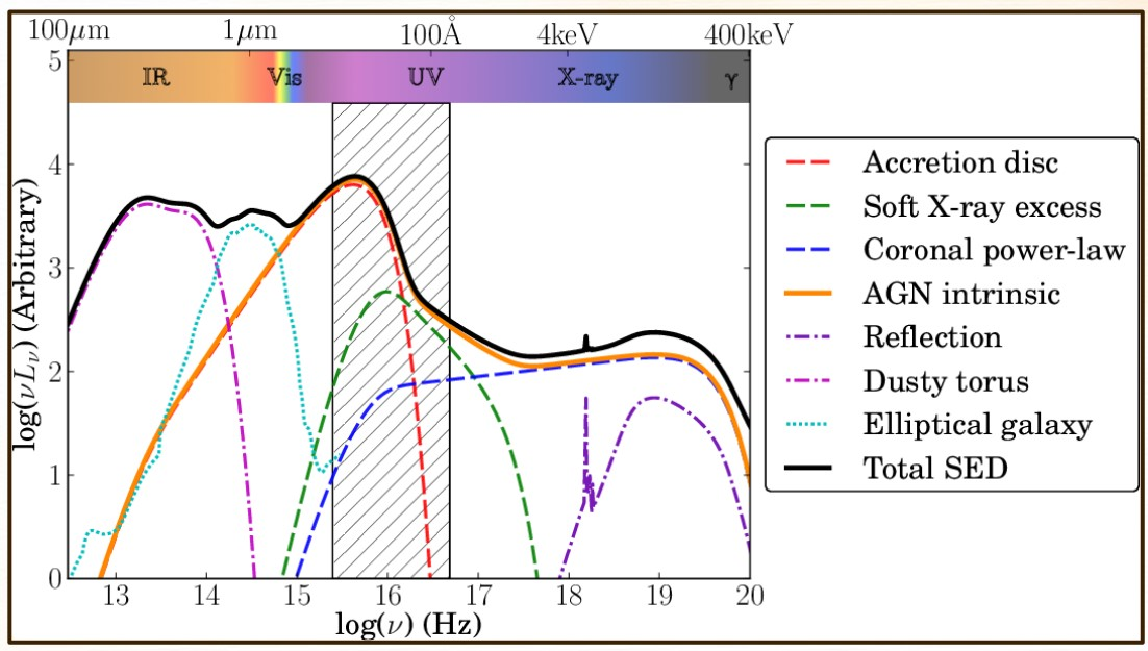}
    \caption{A generic broad-band spectral energy distribution of a Type-1 AGN. The figure is truncated to highlight the various spectral components with the far-infrared to the onset of the $\gamma$-ray region. The shaded region denotes the region affected by absorption due to our Galaxy which dominates primarily the far-UV to soft X-ray region. Figure courtesy: \citet{Collinson_2016PhDT.......352C}.}
    \label{fig:sed}
\end{figure*}

Active galactic nuclei (AGN) are amongst the most luminous cosmic objects to be found in the Universe. They host supermassive black holes (SMBH) at their centers which possess immense gravitational potential. The gravitational force leads to the accretion of matter that in turn leads to heating up,  loss of angular momentum, infall, and formation of an accretion disk \citep{Lyden-Bell_1969Natur.223..690L, ss73}. The radiation emitted due to the heated-up matter from the accretion disk is the primary source of ionization of the gas-rich media in the vicinity of the SMBH. These gas-rich media are photoionized due to the incoming ionizing photons and lead to the production of emission lines which we observe  \citep{schmidt63, greenstein_schmidt64, schmidt_green83, osterbrock_ferland06, netzer2015}. The AGN is made of various interrelated regions that contribute to the observed spectrum that we receive as a spatially unresolved spectrum. 
Apart from the SMBH at the centre and the accretion disk surrounding it,  an X-ray corona is expected to be present where the hot Comptonized X-ray emission originates from. The AGN intrinsic continuum \citep{Collinson_2016PhDT.......352C, kubota18,panda19b, ferland2020} is thus due to the various thermal processes that are marked in Figure \ref{fig:sed}. Primary among them is the thermal continuum from the accretion disk, the X-ray coronal power-law continuum, and the reflection component which is due to the accretion disk reflecting the X-ray photons originally from the corona. In addition to these components, a ``soft X-ray excess'' \citep{arnaud1985}  is partially seen in the spectra of some AGNs. The origin of the soft X-ray excess has been attributed to a warm Comptonizing component and helps to bridge the absorption gap between the UV downturn and the soft X-ray upturn \citep[e.g.,][]{elvis94, laor97, richards2006}, and changes the far-UV and soft X-ray parts of the spectrum.

Emission lines have a range of relative intensities and line widths. Transforming this information into a velocity-space, we then have a range of velocities - for the broad emission lines this velocity is usually $\gtrsim$500-700 \kms{} at low luminosity \citep{zhouetal06,marzianietal09,s11, Rakshit_2020ApJS..249...17R}. 
As these ionized media are bound to the SMBH and they orbit around it in Keplerian trajectories, the exact value of their velocities allows us to infer (relatively) how close to the SMBH they are orbiting. An exact determination of the distances of these line-emitting media can be achieved using the light-echo or the reverberation mapping technique which involves cross-correlating the direct continuum light from the accretion disk and the scattered light from the BLR, both eventually reaching to us \citep{blandford_mckee82, peterson2004, peterson93}. The scattered light from the BLR due to the extra light travel time takes longer to arrive to us relative to the direct continuum light. This allows us to measure the separation of the BLR from the accretion disk. With the collective information of the velocity of the emission line (from single or multi-epoch spectroscopy) and the radial separation of the line-emitting BLR from the ionizing continuum source (or the \rblr{}), one can then use the virial relation to estimate the black hole mass \citep{bentz13, dupu2014, kaspi2000} of the SMBH\footnote{The virial relation also requires a scale-factor (\textit{f}) that contains the information of the geometry of the emitting region.}. Therefore the study of an AGN spectrum should involve the definition of the ionizing continuum and of the intensity ratios and line profile properties of the emission lines. They will in turn provide us with insight on the dynamics, energetics and composition of their immediate surroundings.

\section{Narrow-Line Seyfert 1 Galaxies: drivers of the Eigenvector 1 correlations}

Narrow Line Seyfert Type-1 galaxies (or NLS1s) are a unique class of Type-1 AGNs that are characterized by ``narrower'' broad emission lines. They show FWHM(H$\beta_{\rm broad}$) $\leq$ 2,000 km s$^{-1}$, and the ratio \oiii{}/\hb{} $\lesssim$ 3 \citep{oster85, goodrich1989}. Also, the NLS1s exhibit strong \feii{} emission and the relative strength of the optical \feii{} (within 4434-4684 \AA) to the \hb{}, or \rfe{}, is $\gtrsim$ 1 \citep{sulentic2000, mar18, panda19b}. NLS1s have been used to analyze the \feii{} emission since the late 1970s \citep{phillips1978a} which is among the most noticeable cooling agents of the BLR, emitting about $\sim$25\% of the total energy in the BLR \citep{wills1985}. The \feii{} is a strong contaminant owing to a large number of emission lines and without proper modelling and subtraction, it may lead to a wrong description of the physical conditions in the BLR \citep{verner99, sigut2003, sigut2004, baldwin2004, panda_cafe2}. In addition, the parameter \rfe{} is central to the Eigenvector 1 scheme which consists of the dominant variable in the principal component analysis presented by \cite{borosongreen1992} which is now well understood to be associated with fundamental parameters of the accretion process in the AGNs \citep{sulentic2000, sh14, mar18, panda19b, du2019, martinez-aldama_2021}.

We focus here on NLS1s of relatively high Eddington ratio, selected by the condition \rfe $\gtrsim 1$ \citep{marzianisulentic2014} because they are the class of type-1 AGN most frequently isolated for systematic studies. We stress that a more meaningful approach for low redshift samples would be to impose a limit at FWHM \hb $\approx$ 4000 \kms. This limit defines a Population A of relatively high accretors that obviously incorporates NLSy1s \citep{sulentic2000,mar18}.
NLS1s with high accretion rates are typically shown to have a prominent soft-X-ray excess \citep{arnaud1985} that can be clearly seen in their broadband SED \citep{jin12a, jin12b, kubota18, ferland2020}.

\section{Shorter BLR time-lags and higher accretion rates in NLS1s}

The recent reverberation mapping results have led to populate the empirical \rblr{}-\lopt{} observational space \citep{kaspi2000,bentz13} and taking the total count over 100, especially the sources monitored under the SEAMBH project (Super-Eddington Accreting Massive Black Holes, \citealt{dupu2014, Wang2014_seambh, hu15, du2015, dupu2016a, dupu2018}), and from the SDSS-RM campaigns \cite{grier17, Shen_2019ApJS..241...34S}. But this has introduced us to a new challenge - the inherent dispersion in the \rblr{}-\lopt{} relation after the introduction of these new sources (see Figure 1 in \citealt{Panda_Marziani_2022arXiv221015041P}). The sources that eventually led to the increase in the scatter in the relation, show an interesting trend with the Eddington ratio - the larger the dispersion of a source from the empirical \rblr{}-\lopt{} relation, the higher its Eddington ratio.  \citet{martinez-aldama2019}  found that this dispersion can be accounted for in the standard \rblr{}-\lopt{} relation with an added dependence on the Eddington ratio (\lbol{}/\ledd{}). \citet{du2019} exploited this further and realized that with an additional correction term, the relation can be reverted back to the original relation with a slope $\sim$0.5. This additional correction term is the strength of the \feii{} emission in the optical region - \rfe{}, which has been shown in earlier studies to be a reliable observational proxy for the Eddington ratio \citep{sulentic2000, sh14, mar18, panda19b, du2019, martinez-aldama_2021}. Thus, the introduction of \rfe{} in the \rblr{}-\lopt{} relation, is able to account for the shorter time-lags and hence, smaller \rblr{} sizes for sources that show strong \feii{} emission.

\section{Anisotropic emission from the disk - the missing ingredient?}

\begin{figure*}[!htb]
    \centering
    \includegraphics[width=\textwidth]{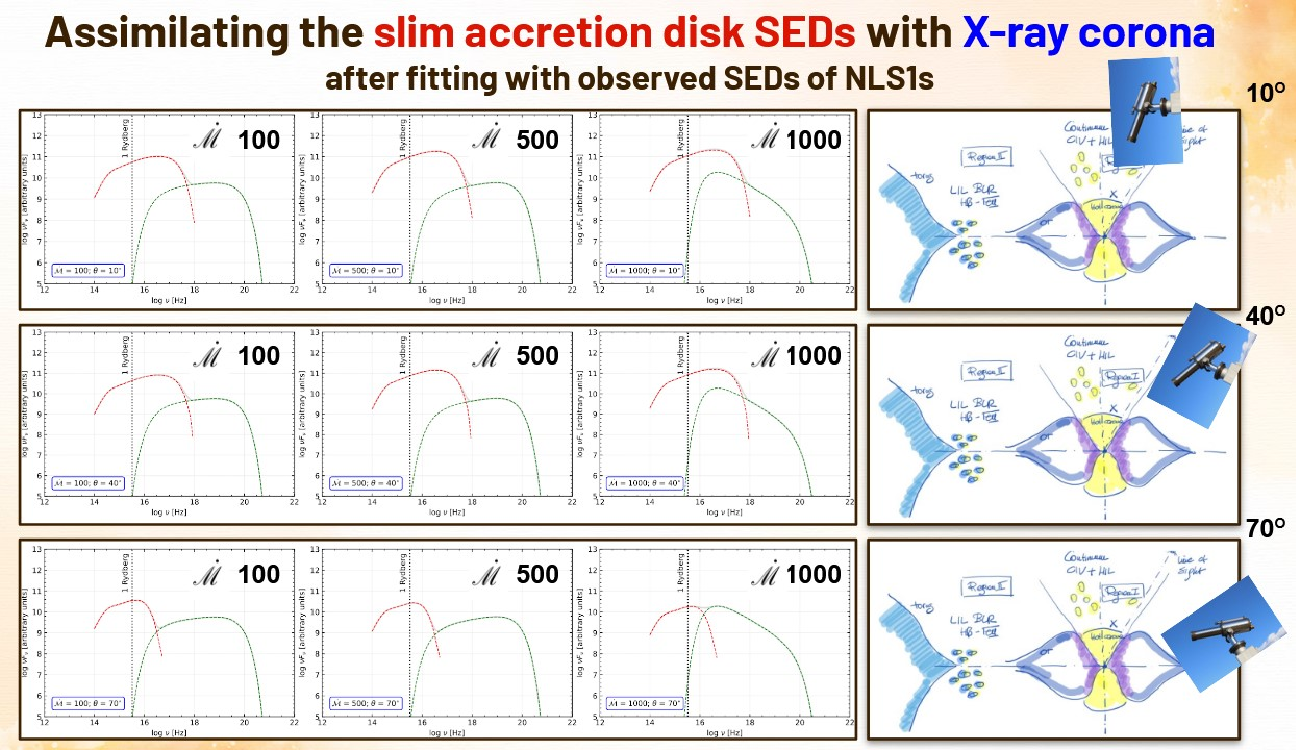}
    \caption{A montage of slim accretion disk plus hot coronal component composite SEDs. The slim-disk models \citep{wang14} are shown for three representative dimensionless accretion rates (\mdot{}): 100 (left panels), 500 (middle panels), and 1000 (right panels). Three cases of viewing angle (with respect to the distant observer as shown using the telescope on each of the panels in the extreme right) are shown, i.e., \textit{i} = 10$^{\circ}$ (upper panels), 40$^{\circ}$ (middle panels), and 70$^{\circ}$ (bottom panels). These models are made for a representative black hole mass, \mbh{} = 10$^8$ \msun{}. The X-ray corona component is adjusted based on the normalization with respect to the slim disk, i.e., using the spectral index $\alpha_{\rm ox}$ which is obtained by fitting the three SEDs at 10$^{\circ}$ (upper panels) to their respective counterparts from observations arranged with respect to their \mdot{} values \citep[see Figure 1 in][]{ferland2020}. A schematic view of the inner sub-parsec region around the SMBH for a high accreting AGN is shown on the extreme right. Abridged version from \cite{wang14}; not drawn to scale.}
    \label{fig:slim-disks}
\end{figure*}

Accounting for these shorter time lags for the observed AGNs needs advancement in our photoionization modelling setup, to account for the Eddington ratio dependence on the accretion disk emission and the changes it causes in the overall disk vertical structure, especially the regions very close to the last stable orbit around an accreting SMBH. \citet{wang14} derived the analytical solutions (steady-state) for the structure of ``slim'' accretion disks from sub-Eddington accretion rates to extremely high, super-Eddington rates. They notice the appearance of a funnel-like structure very close - few gravitational radii from the SMBH (see Figure 3 in their paper) and attribute this change in the accretion disk structure to the high accretion rates that are realized from the solutions of the slim accretion disks wherein the structure of the geometrically thin, optically thick accretion disk as per the \cite{ss73} prescription does not hold \citep[][]{abramowicz88, sadowski2011, wang14}. We show an illustration of this scenario in the right panel of Figure \ref{fig:slim-disks}. Such modiﬁcations to the disk structure strongly affect the overall anisotropic emission of ionizing photons from the disk in addition to just inclination effects that arise due to the axisymmetric nature of these systems. Therefore, a viewing-angle-dependent anisotropy needs to be accounted for in the modelling at high accretion rates. If the gas clouds emitting the low-ionization lines are located not far from the disk plane, the lower disk luminosity (with respect to the one seen by an observer whose line-of-sight is almost coaxial with the disk axis) leads to the shrinking in the position of the line emitting BLR and brings the modelled location in agreement to the observed estimates from the reverberation mapping campaigns \citep[see][for more details]{panda_cafe2}.

In our setup, we construct a database of composite SEDs which comprise two primary components - a slim accretion disk and a hot corona (shown using red and green curves in each of the panels of Figure \ref{fig:slim-disks}). The normalization of these two components is characterized by the spectral index ($\alpha_{\rm ox}$) - which is the ratio of the flux at 2 keV (from the hot corona) and flux at 2500\AA\ (from the accretion disk). The exact estimate of this parameter and the overall slope of the two spectral components in our SEDs is set by matching the modelled composite SEDs viewed at 10$^{\circ}$ with observed SEDs from \citet{ferland2020}. For each observed SED case, we find the best representative composite SED from our models. The upper panels in Figure \ref{fig:slim-disks} show these best-fit modelled SEDs as a function of increasing dimensionless accretion rate (\mdot{}, see \citet{Wang2014_seambh, Panda_2022FrASS...950409P} for an overview). In addition, we extend the modelled SEDs as a function of angle in order to quantify the effect of the viewing angle to the distant observer. In principle, we assume that the hot corona comes from an isotropically radiating source and hence is not affected by the change in the viewing angle, although the accretion disk component does change with the viewing angle. The illustration on the right panels in Figure \ref{fig:sed} highlights this aspect of the anisotropy. The funnel-like structure leads to a preferential direction of the continuum radiation coming from the accretion disk, especially the photons that originate very close to the SMBH. 

The high-ionization part of the BLR contains a set of lines that include \lya{}, \ciii{}, \civ{}, \hei{}, \heii{}, and \nv{} emitted by a highly ionized region that has a relatively low density ($\lesssim$10$^{10}$ \cc{}). Whereas, the low-ionization part of the BLR includes the bulk of the Balmer lines, \mgii{}, \feii{}, \oi{} and \caii{}, emitted by a mildly ionized medium having a much higher density ($\gtrsim$10$^{10}$ \cc{}).\footnote{Reverberation mapping estimates have shown that the two emitting regions have different distances from the SMBH - the high-ionization lines (HILs) are produced much closer to the SMBH while the low-ionization counterparts originate from a region that is relatively more distant \citep{grier17,Lira_2018ApJ...865...56L,Homayouni_2020ApJ...901...55H,Kaspi_2021ApJ...915..129K}. This result is in apparent contradiction with the scenario we suggest. This is not really the case: these results refer to Population B sources \citep{sulentic2000}, which are accreting at modest rates.} Recent studies \citep{panda_cafe2, Panda_2022FrASS...950409P} have shown that the optimal emission from the LIL-part of the BLR requires a much lower radiation field, i.e., lower than expected photon flux that is incident on these emitting regions which is then able to recover the line equivalent widths (EWs) for the LILs. Also, this prescription has qualitatively shown that the location estimated for these emitting regions is then in agreement with the estimates derived from the reverberation mapping of the NLS1s. The overall effect suggests that the fraction of the photon flux originating from the accretion disk and corona that eventually is responsible for the line emission, in the LIL-part of the BLR, is a very small ($\sim$1-10\%). Although this is a step in the right direction, there is yet to have a full quantitative picture of the overall emission from the low- and high-ionization line emitting regions. Here, we further our investigation by accounting for the changes in the disk structure that can eventually lead to the development of a geometrically thick, funnel-like structure within a few 100 gravitational radii that can block a significant part of the incident radiation directed towards the LIL part. This funnel then leads to a preferential emission in a cone about the spin axis of the SMBH where the HILs can be produced.

\begin{figure}[!htb]
    \centering
    \includegraphics[width=0.5\textwidth]{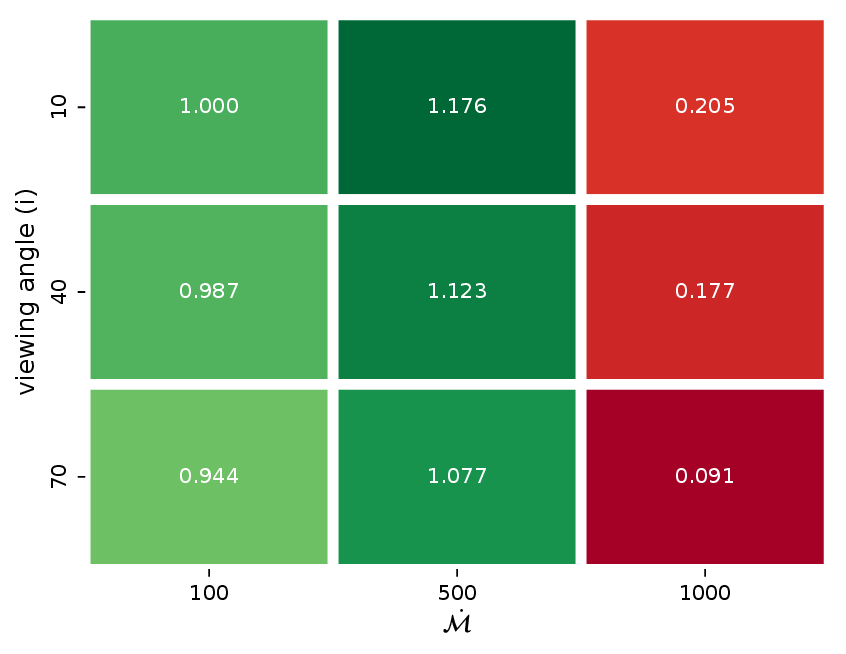}
    \caption{Relative area under the composite SEDs shown in the panels of Figure \ref{fig:slim-disks} (wrt the case with \textit{i} = 10$^{\circ}$, \mdot{} = 100)}
    \label{fig:ratio}
\end{figure}

Coming back to Figure \ref{fig:slim-disks}, we estimate the total ionizing flux in each of these composite SEDs shown as a function of the \mdot{} and viewing angle. The heatmap shown in Figure \ref{fig:ratio} highlights this aspect of the composite SEDs. The values in each grid mark the ratio of the photon flux (above 1 Rydberg) for a given case (\mdot{} and viewing angle) with respect to the base case, i.e., the composite SED at \mdot{}=100 and viewing angle=10$^{\circ}$ (upper left panel in Figure \ref{fig:slim-disks}). As we can gauge from this heatmap, for any given value of \mdot{} (100, 500, and 1000), the amount of ionizing photon flux available decreases with increasing viewing angle. Whereas, the trend with respect to increasing \mdot{} for a given viewing angle first rises (going from 100 to 500) and then falls off drastically (from 500 to 1000). The highest available ionizing photon flux is obtained for the case with \mdot{} = 500 viewed at 10$^{\circ}$. On the other hand, the lowest ionizing photon flux is recovered for the case with \mdot{} = 1000 viewed at 70$^{\circ}$.

\section{Concluding Remarks}

The inclusion of the anisotropic emission from a slim disk is therefore going qualitatively in the right direction: the HILs may be emitted in a funnel or at any rate in a region fully exposed to the ionizing continuum. On the converse, LILs might be exposed to a fainter continuum that requires a closer distance in order to compensate for the lower photon flux. At present, this is our working hypothesis. A more detailed analysis of the results from our photoionization modelling will be presented in a forthcoming work.

\begin{acknowledgements} SP acknowledges the Conselho Nacional de Desenvolvimento Científico e Tecnológico (CNPq) Fellowships 164753/2020-6 and 300936/2023-0. We are grateful to Prof. Jian-Min Wang for providing his slim disk SED models, and to Prof. Bo\.zena Czerny for fruitful discussions.  \end{acknowledgements} 

\bibliographystyle{mnras}
\bibliography{references}

\end{document}